\documentclass[reprint,onecolumn,amsmath,amssymb,aps,prb,showpacs]{revtex4-2}

\usepackage{amsfonts}
\usepackage{latexsym}
\usepackage{graphicx}
\usepackage[usenames]{color}
\usepackage[utf8]{inputenc}
\usepackage{hyperref}
\usepackage{epstopdf}

\newcommand{\be}{\begin{equation}}
\newcommand{\ee}{\end{equation}}
\newcommand{\nn}{\nonumber}
\newcommand{\Fref}[1]{Fig.~\ref{#1}}
\newcommand{\Eqref}[1]{Eq.~(\ref{#1})}

\begin{document}
\title{EPO10, 7 December 2023, Skopje\\
Studying of well-known and widely used technical patent.
The principle of operation of auto-zero and chopper-stabilized DC amplifiers
}

\author{Todor~M.~Mishonov}
\email{mishonov@gmail.com}

\author{Albert~M.~Varonov}
\email{akofmg@gmail.com}

\author{Peter~V.~Peshev, Vasil~G.~Yordanov}
\affiliation{Physics Faculty, St.~Clement of Ohrid University at Sofia, \\
5 James Bourchier Blvd., BG-1164 Sofia, Bulgaria}

\author{Stojan~G.~Manolev}
\affiliation{Middle School ``Goce Delchev'',\\
Purvomaiska str. 3, MKD-2460 Valandovo}

\author{Riste Popeski-Dimovski}
\email{ristepd@gmail.com}
\affiliation{Institute of Physics, Faculty of Natural Sciences and Mathematics, 
``Ss. Cyril and Methodius'' University, Skopje, R.~N.~Macedonia}

\begin{abstract}
The problem given to the 10 Experimental-Physics Olympiad (EPO10)
7 December 2023, Skopje, 11:00--15:00. 
The Olympiad is organized by 
Society of Physicists Macedonia.
and the Sofia Branch of the Union of Physicists in Bulgaria.
The experimental set-up illustrates the principle 
of chopper stabilized DC amplifiers described in 
the patent by 
Edwin Goldberg and Jules Lehmann, 
\emph{Stabilized direct current amplifier},
U.S.~Patent~2,684,999 (1949).
Now this principle is implemented in many contemporary
operational amplifiers.
In such a way this student problem is related to the
operation of widely used technical device
and many participants successfully solved the problem.
\end{abstract}

\date{29 Dec 2023, 15:30}

\maketitle

\section{EPO10 Problem}
\begin{figure}[ht]
\includegraphics[scale=0.7]{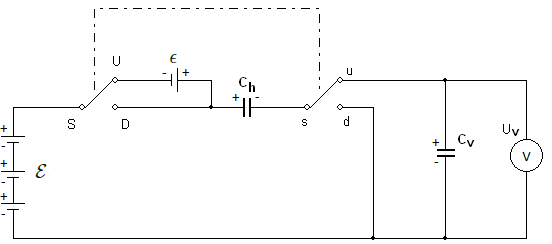}
\caption{Circuit of charging of capacitors with paired switches.}
\label{circuit}
\end{figure}

\begin{figure}[ht]
\begin{minipage}[t]{0.4\linewidth}
\centering
  \includegraphics[scale=0.15]{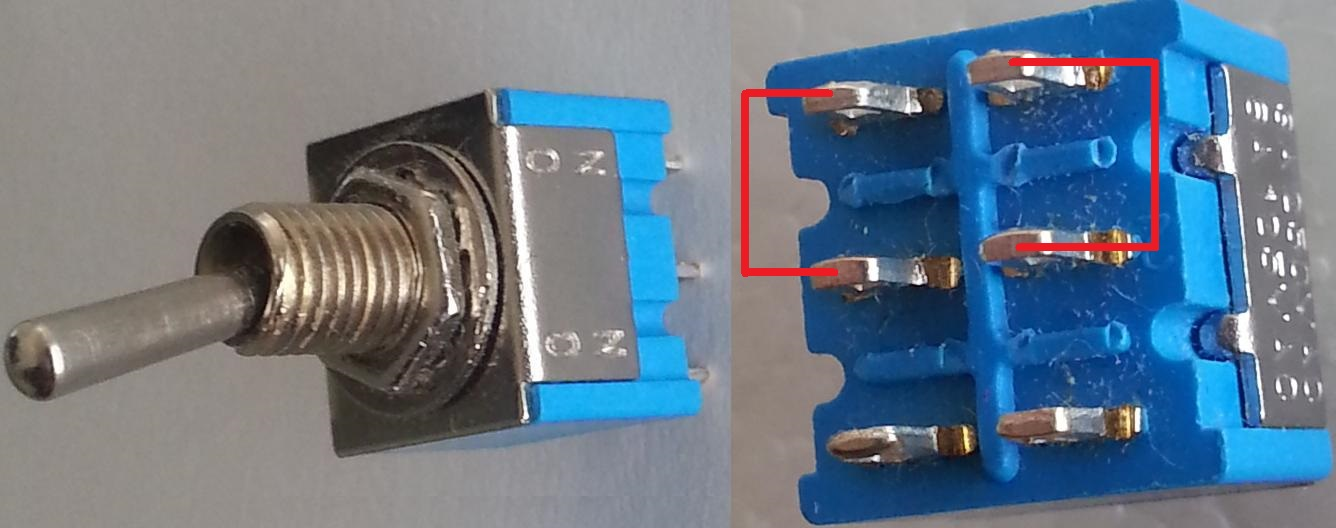}
  \caption{Double switch in up position.}
  \label{switch_up_wired}
 \end{minipage}
\begin{minipage}[t]{0.4\linewidth}
  \centering
  \includegraphics[scale=0.15]{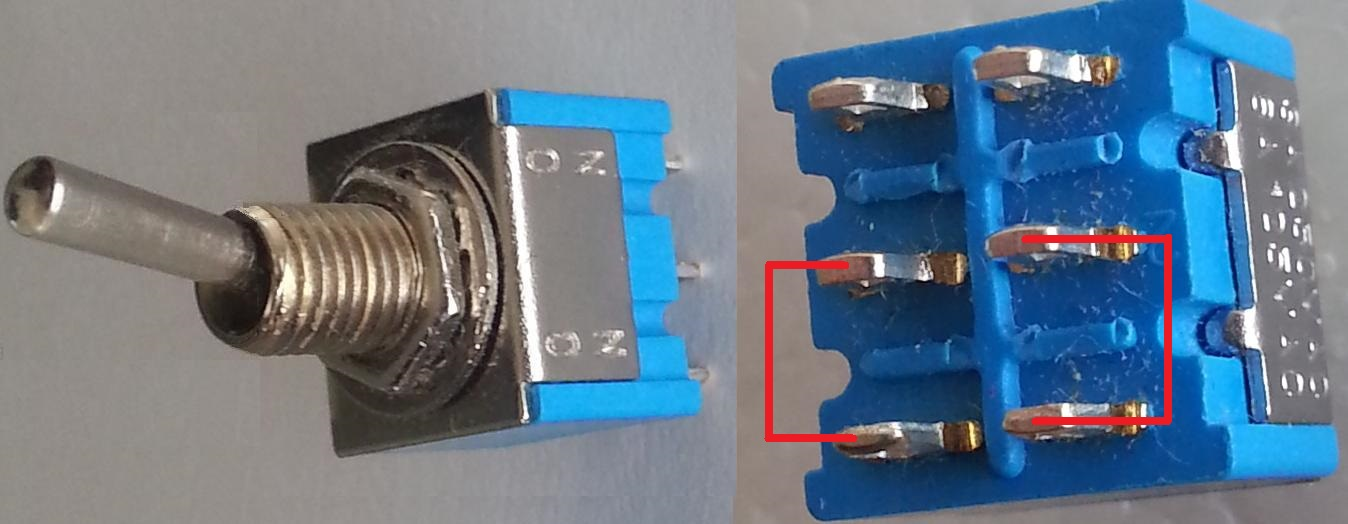}
  \caption{Double switch in down position.}
 \label{switch_down_wired}
\end{minipage}
\end{figure}

On the scheme (circuit) shown in \Fref{circuit} both switches are paired (also called bipolar on/off switch or double switch)
and both can be simultaneously up (marked with ``U'' and ``u''), see \Fref{switch_up_wired} or down (marked with ``D'' and ``d''), see \Fref{switch_down_wired}.
The capacitors with capacitance $C_\mathrm{h}$ and $C_\mathrm{v}$ are charged, by switching, from the both sources of electrical voltage $\mathcal{E}$ and $\epsilon$.

\begin{itemize}

\item Measure with voltmeter the voltage $U_\mathrm{v}$ of the capacitor $C_\mathrm{v}$ after performing big enough number of switchings of the paired switches.

\item How the voltage $U_\mathrm{v}$ depends on $\mathcal{E}$,  $\epsilon$, $C_\mathrm{h}$ and $C_\mathrm{v}$ and where this task could find technical application?

\item Measure the voltage $U_\mathrm{h}$ at capacitor $C_\mathrm{h}$ after performing big number fast switching.

\end{itemize}

If you can solve the task in your own way, you could start working without reading the instructions. Students who can not perform the experiment 
may try to predict the result of the experiment without measurements 
- this is also accepted as a solution.

\section{Guidelines for accomplishment of the tasks}
If you do not know how to start, follow the instructions step by step, 
corresponding to the numerated tasks.
\textbf{Attention: do NOT give the batteries on short-circuit!} 
If you try to measure the voltage of a battery, 
and the multimeter is turned on as an ampere-meter
you can burn out the ampere-meter mode.
\textit{All measurements with the multi-meters should be performed using their most accurate possible range depending on the task.}

Different tasks with increasing difficulty 
are described in the sections corresponding to each age category.
These tasks lead to the entire solution of the problem of the EPO10.
Try to solve as much tasks as possible 
without paying attention to the age category.
The age-category serves for evaluation purpose mainly.
Please, keep the humanitarian text as short as possible.

\begin{figure}[ht]
\includegraphics[scale=0.1]{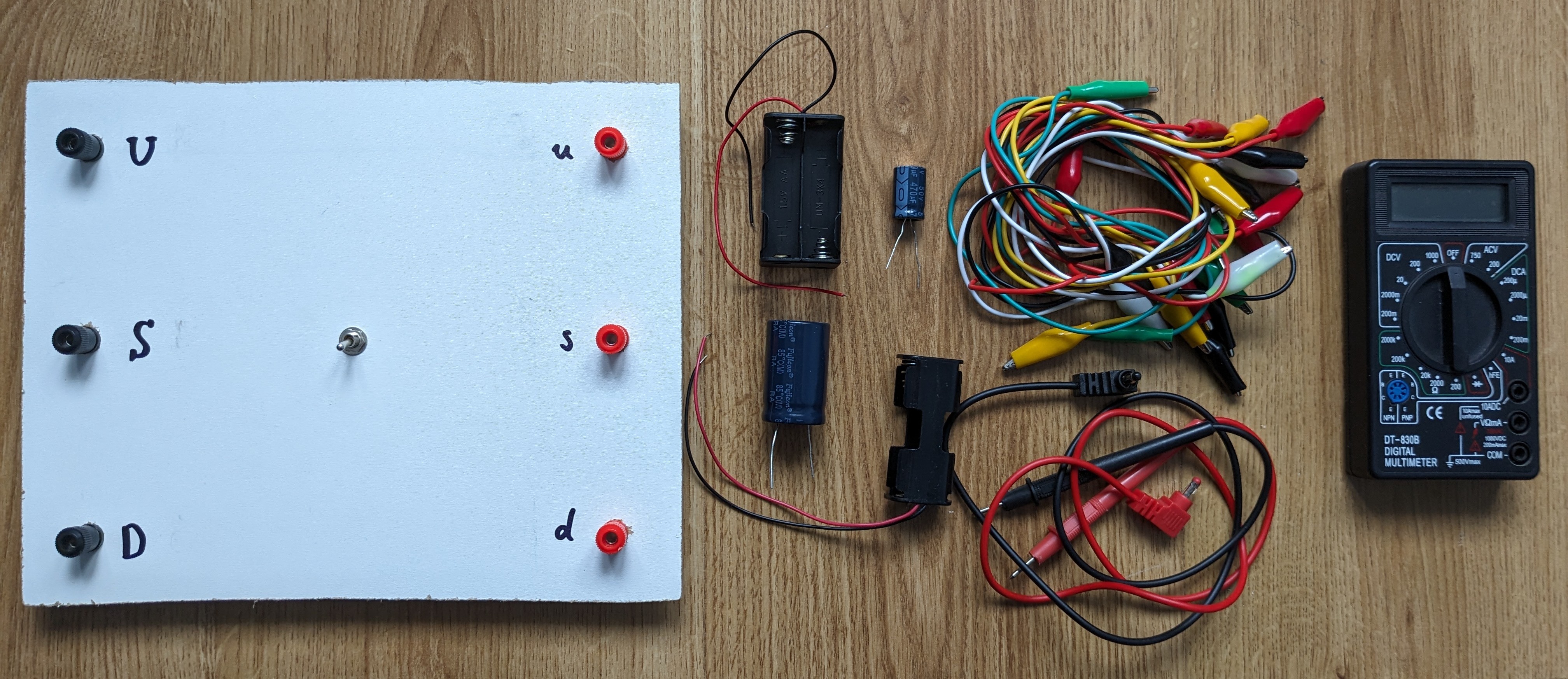}
\caption{A photo of the entire experimental set-up.
Only one 1 multi-meter is shown (right), which is the necessary minimum for performing the experiment.
One can see a wooden panel (left) with the double switch in its center, 2 different electrolytic capacitors, holders for 4 and 2 type AA batteries~ (middle), 12 crocodile cables and a pair of probes for the multi-meter~(right).
}
\label{setup-photo}
\end{figure}

In \Fref{setup-photo}, a photo of the experimental set-up is given.
This picture might be a little bit different if additional pair of multimeter probes ending with crocodile clips are given to be used for connecting the sockets of the panel labeled with ``u'', ``s'' and ``d'', both lowercase and uppercase.

\section{Tasks S. Getting to know the voltage source}
\label{Sec:beginning}

\begin{enumerate}

\item 
Switch the multimeter as a voltmeter.
Plug the test probes in the multimeter. 
Measure and write down the voltage of all 4 given 1.5~V type AA batteries.
Use the maximal accuracy range of the voltmeter.

\item Connect two batteries with each other when the plus of the first is connected with the minus of the other, this is a series of consecutive connections, 
see~\Fref{battery_plus_minus}. 
Measure the voltage and compare it with the sum of the voltages of the two batteries. 
Now connect the both pluses of the batteries and measure the voltage between the two minuses, see \Fref{battery_plus_plus}.
What is the difference between the calculated and the measured values?
\begin{figure}[ht]
\begin{minipage}[t]{0.45\linewidth}
   \includegraphics[scale=0.6]{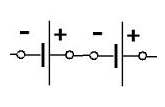}
   \caption{Connecting the batteries in series, 
   the plus of one of the batteries is connected 
   with the minus of the other battery.}
   \label{battery_plus_minus}
\end{minipage}
\qquad
\begin{minipage}[t]{0.45\linewidth}
   \includegraphics[scale=0.6]{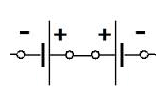}
   \caption{Connecting the batteries in series, 
   the plus of one of the batteries is connected to the plus of the other battery.}
   \label{battery_plus_plus}
\end{minipage}
\end{figure}

\item
Put a single battery in the 2-battery holder and use crocodile-clips cable to short-circuit the remaining empty place for the missing battery, see \Fref{2-AA}, and write down the voltage $\epsilon$, it should be slightly above 1.5~V.
\begin{figure}[ht]
\begin{minipage}[t]{0.4\linewidth}
  \includegraphics[scale=0.4]{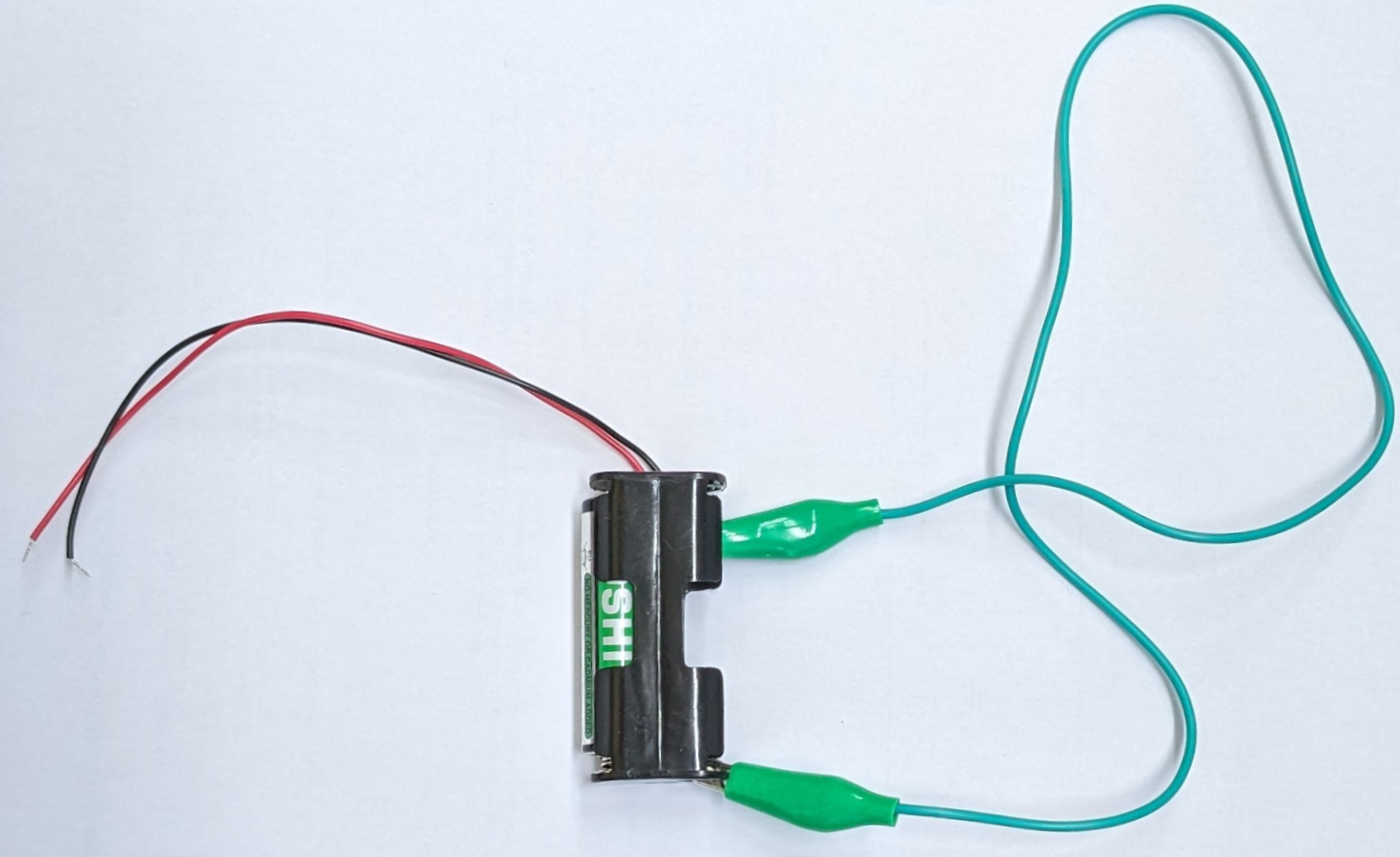}
  \caption{One battery in 2-battery holder.
  One crocodile cable gives a short circuit to the missing battery.
  In such a way we obtain a voltage source with voltage $\mathcal E_1$.}
  \label{2-AA}
 \end{minipage}
 \qquad
\begin{minipage}[t]{0.55\linewidth}
  \includegraphics[scale=0.58]{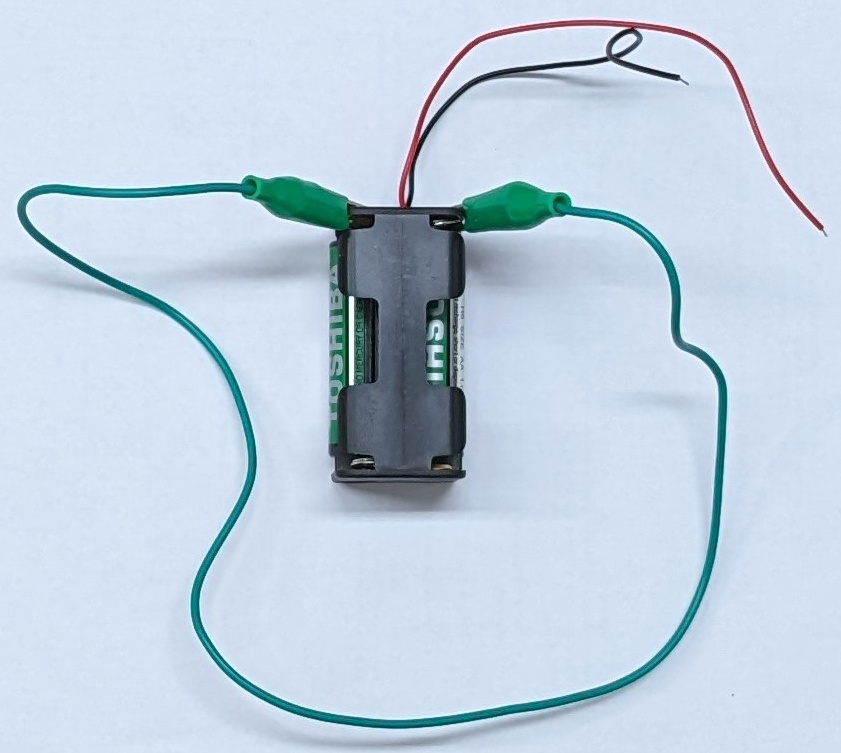}
  \caption{Two batteries in a 4-battery holder.
  The crocodile cable gives short circuit to the places of two missing   
   batteries and we have voltage source with voltage 
   $\mathcal E_1+\mathcal E_2$.
   Analogously we can obtain voltage 
   $\mathcal E_1+\mathcal E_2+\mathcal E_3$ 
   using 3 batteries, or $\mathcal E_1$ using only one battery.
   }
 \label{4-AA}
\end{minipage}
\end{figure}

\item
Analogously, 
put three batteries in the 4-battery holder and using a crocodile-clips cable to short-circuit the empty place for the missing battery measure the voltage $\mathcal E$, see \Fref{4-AA}; 
it should be slightly above 4.5~V.
Now you have voltage sources for the electric scheme (circuit).

\item Use the 1.5~V voltage source, crocodile cable and charge 
470~$\mu\mathrm{F}$ capacitor.
Minus probe of the battery must be connected to the minus lead of the capacitor.
Minus lead (pin) of the electrolytic capacitor is marked by a series of ``$-$'' signs on the capacitor and it is shorter than the ``$+$'' lead (pin).
For the light emitting diodes (LED) is used the same convention.
Use the voltmeter to check whether the battery is properly connected to the capacitor.
Now the capacitor is charged.

\section{Tasks M. Charging and discharging of capacitors}
\label{sec:M}

\item 
Now prepare to write down the voltage $U$ 
as function of time $t$ as it is suggested on 
Table~\ref{example_table_C_discharge}. 
Remove the voltage source and write down the voltage $U_i$ 
and time $t_i$
every 30 seconds during 10 minutes.
Those are 21 experimental points.
Use the maximal possible accuracy range of 2000~mV
when the capacitor is charged by the 1.5~V battery.
Leave space for an extra column which will be filled in later.
\begin{table}[ht]
\begin{tabular}{ r  r  r  r}
\tableline \hline
i & $\qquad t_\mathrm{i}$ [s] & $\qquad U_\mathrm{i}$ [mV] & 
$\qquad  -\ln(U_\mathrm{i}/U_0)$\\
\tableline 
0 & \dots &  \dots & \dots\\
1 & \dots & \dots & \dots \\
2 & &  \\
3 & &  \\
\vdots & \dots  &  \dots & \dots  \\
\tableline \hline
\end{tabular}
\caption{The voltage of the capacitor as a function of time. 
In the last column, 
$U_0$ is the voltage of the first measurement 
labeled by the index i=0, or $U_0 \equiv U_\mathrm{i=0}\equiv U(t=0)$.
Do not draw vertical lines in tables.}
\label{example_table_C_discharge}
\end{table}

\item Now take a graph paper and present graphically the results from the table. 
Draw smooth curve, which passes close to the measured data points.
Choose the most optimal scaling so that the graph to be as large 
and as clear as possible.

\item Extend Table~\ref{example_table_C_discharge} 
with an extra column (column~4)                        
for which calculate and fill in $-\ln[U_\mathrm{i}/U_0]$, 
where $U_0$ is your first voltage measurement 
at $t=0$ and $\mathrm{i}=0, 1, 2, \dots$.

\item
Present graphically $-\ln \left[U(t)/U(0)\right]$ 
(Table~\ref{example_table_C_discharge}, column 4)
versus $t$ and draw a straight line through the cloud of points.
What quantity reveals the slope of the straight line in this plot?

\item
Choose two points in the approximating line and determine
the difference between the abscissa $\Delta t$
and ordinate $-\Delta\ln \left[U(t)/U(0)\right]$,
i.e. the slope of the line
\be
a = \frac{-\Delta\ln \left[U(t)/U(0)\right]}{\Delta t} \equiv
 \frac{-(\ln \left[U(t)/U(0)\right]_\mathrm{B}
 -\ln \left[U(t)/U(0)\right]_\mathrm{A})}{t_\mathrm{B}-t_\mathrm{A}}
=\frac1{\tau}
\nn
\ee
for two points named A and B, for instance.
A simplification is possible: 
in our case $t_\mathrm{A}=0$
and
$-\ln \left[U(t)/U(0)\right]_\mathrm{A}=0$, as well.

\item
Calculate the quantity 
$\tau \equiv 1/a = \Delta t/\left(-\Delta\ln \left[U(t)/U(0)\right]\right).$
What is the dimension of this quantity and why?

\item Switch the other multimeter as an Ohm-meter and measure the 
internal resistance $R$ of the used up to now voltmeter and write it down.

\item Calculate the capacitance of the capacitor $C$ from $\tau = R C$ and write it down.

\item 
Using a crocodile cable connect the probes of the capacitor.
This short circuit should be at least 30 seconds
and remove the cable. 
What voltage shows the voltmeter;
write down this small voltage.
Now we can say that the capacitor is discharged.

\item If one of your multi-meters has the option to measure a capacitance,
measure $C$ with it. 
If not, take the nominal value of $C$ indicated on the body of the used up to now capacitor.

\item Compare both results for $C$ and present their difference in \%.
Which one is more reliable according to you and why 
(having in mind that the nominal value of $C$ is accurate within 20\%)?

\section{Tasks L. The experiment with the given circuit}
\label{sec-L}

\item Connect the circuit shown in \Fref{circuit} using the given experimental set:
$C_\mathrm{h}=470\;\mu\mathrm{F}$, $C_\mathrm{v}=4700\;\mu\mathrm{F}$, $\epsilon\approx1.5\;\mathrm{V}$ and $\mathcal{E}\approx4.5\;\mathrm{V}$ and the wooden panel with the double switch and the sockets 
labeled ``u'', ``s'' and ``d'' (both lowercase and uppercase).
The number of cables is limited,
use the insulating head of the sockets
in order to screw one probe (pin) of the capacitors 
to the metal ring as it is shown in \Fref{socket_with_capacitor},
sparing one cable for this connection.
\begin{figure}[ht]
\includegraphics[scale=1.0]{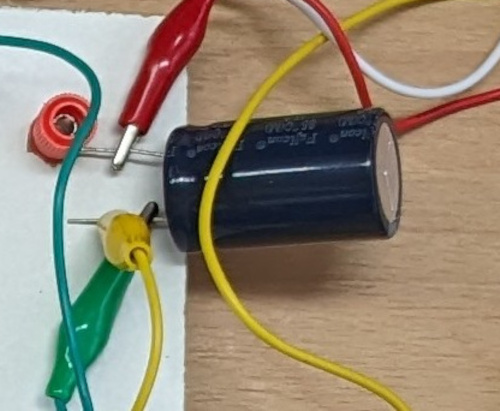}
\caption{Longer plus lead (pin) of the electrolytic capacitor
is screwed by the insulating cap to the metal ring of the socket.
Crocodiles are connected to the capacitor pins.}
\label{socket_with_capacitor}
\end{figure}

\item After connecting the circuit, check the polarity of the electrolytic capacitors. 
The sign 'plus' is not shown on the capacitor. 
The sign 'minus' is marked with one or several signs (-, - -, - - -). 
In case of inappropriate connection of the polarity, 
the electrolytic capacitors are discharged very quickly
and the measurement cannot be performed 
with maximal accuracy. 

\item Measure the voltage $U_\mathrm{v}$ of the capacitor 
$C_\mathrm{v}$ after performing large enough number 
of fast switching of the double/paired switch and write it down.
Continuously switch the switch 
and observe the readings (testimonies) of the voltmeter.

\item Repeat the same experiment measuring the voltage 
of the other capacitor $U_\mathrm{h}$.
Having 2 multi-meters you can measure those voltages simultaneously.

\item Make a new table with 5 columns
as it is shown on Table~\ref{example_table}:
1) the number of the measurement, 
2) voltage at position $\mathcal E$,
3) voltage at position $\epsilon$,
4) voltage at capacitor $U_\mathrm{v}$,
5) voltage at capacitor $U_\mathrm{h}$.
Write down the results of your last measurements in the table.
\begin{table}[ht]
\begin{tabular}{ c  c  c  c  c}
\tableline \tableline \tableline
\# &$\quad \mathcal{E}$ (V) & $\quad \epsilon$ (V) 
& $\quad U_\mathrm{v}$ (V) & $\quad U_\mathrm{h} (V) $ \\
\tableline \hline
1&4.5 & 1.5 & \dots & \dots \\
2&3.0 & 1.5 & \dots & \dots \\
3&1.5 & 1.5 & \dots & \dots \\
4&0.0 & 1.5 & \dots & \dots \\
\tableline \hline
\end{tabular}
\caption {Example for presenting the experimental data; 
you have to explore all 4 combinations.
Here, the values for $\mathcal{E}$ and $\epsilon$ are examples.
You have to fill in your measured values during the experiment.
Usually, new AA 1.5~V batteries have typical voltage $\gtrsim 1.6$~V.
Do not draw vertical lines in tables.}
\label{example_table}
\end{table}

\item 
Repeat the measurement of $U_\mathrm{v}$ 
and $U_\mathrm{h}$ after performing large enough number of fast 
switchings of the paired switch (circuit breaker) 
using different $\mathcal{E}$ sources 
with different voltages: 
$0.0$~V, $1.5$~V,  $3.0$~V and $4.5\;\mathrm{V}$
(or no battery, 1~battery, 2~batteries and 3~batteries). 
For this purpose you have to use 4-battery holder
and a crocodile cable to short-circuit the remaining empty place(s) for battery as it is shown in \Fref{4-AA}.
Present the data in a table, 
as shown in Table~\ref{example_table}. 
Zero voltage is obtained, when you remove all batteries 
and replace them with a conductor (piece of wire or a cable in the given set).

\item Analyze the table with the experimental data and try to determine a common rule for calculation of $U_\mathrm{v}$ by using  the parameters of the circuit $\mathcal{E}$,  $\epsilon$, $U_\mathrm{h}$ and $U_\mathrm{v}$. 
Is it possible to derive this common rule 
by theoretical consideration?
``We discover with intuition -- we prove with logic''. 

\section{Tasks L and XL. Possible practical application of the solved problem?}
\item At the end, \textbf{only after the experimental solution of the whole ``L'' problem from Sec.~\ref{sec-L}},
you may consider where this problem could find technical application.
This is the most difficult task requiring understanding and thinking.

\section{Tasks L and XL. Theoretical problems related to the experiment}

\item The charged capacitor is discharged through the internal resistance $R$
of the voltmeter, used in the experiment in Sec.~\ref{sec:M}.
Derive the time dependence of the voltage of the 
slowly discharging capacitor $U(t)=U(0)\exp(-t/\tau)$
where time constant $\tau=RC.$
Take the logarithm and derive
$-\ln \left(U(t)/U(0)\right)=t/\tau$.
Describe how this linear dependence can be used to determine 
the capacitance $C$.
Derive the formula $\tau=RC=t/\ln\left(U(0)/U(t)\right)$.

\item
Suggest an easier and quicker method for the determination of $\tau$ 
than the one proposed here.
Use it to calculate $\tau$ again and compare it with the obtained value from the experiment in Sec.~\ref{sec:M}, present the difference in \%.
Make short comment about the reliability of the suggested method in terms of accuracy versus necessary effort.
Physics does not comprise endless calculations and measurements,
sometimes not so accurate estimation is preferable.

\item To analyze the circuit (scheme) you have to draw it twice. 
Once with switches in the up position 
and once again 
with switches in the down position. 
After that redraw the schemes and omit the inessential details, 
for example, 
the loose end of the conductors. 
Can you calculate the value of $U_\mathrm{v}$ after performing
a large number of switchings? 
Is there an agreement between the calculation and the measurement?

\end{enumerate}

\section{Afterglow of ``EPO bio-field''}

Determine the capacitance $\mathcal C$ of the bigger capacitor 
with nominal capacitance
$4700\,\mu$F.
Discharging the capacitor in this case is very slow 
$U(t)=U(0)-\Delta U(t)$
and for 10 minutes $\Delta U(t)\ll U(0)$.
In this case, 
$\ln\left(U(0)/U(t)\right)\approx \Delta U(t)/U(0)\ll1$
and 
$\mathcal C\approx (t/R)\left[ U(0)/\Delta U(t)\right]$.
Charge the capacitor for 1 minute and measure at the 
end the voltage $U(0)$.
Switch off the voltage source, this is the moment $t=0$.
After five minutes ($t=300\,$s) 
measure the decrease voltage of the capacitor $\Delta U(t)$.
Can you invent the circuit (scheme) for which
$\Delta U(t)$ is measured with maximal 
sensitivity range of 200~mV of the multi-meter.
Evaluate in percentage \%  the accuracy 
$100\times|\mathcal C/(4.7\,\mathrm{mF})-1|$
of the difference between the measured and the nominal voltage.
Send your solution to epo.mkd@gmail.com within a week.

One of the purposes of the Olympiad is to enhance the skills of participants
seeking a carrier in science or engineering.
That is why, we are giving the experimental set-up as a gift 
to all participants so that every 
one of them 
can repeat at home 
the experiment and hopefully 
will be able to reach the level of the Olympiad champion.

The solution of the problem will be published in arXiv (\url{https://arxiv.org/})
within one month.


\clearpage

\appendix
\section{Solution of the EPO10 Problem}

\subsection{The answer of the problem of this EPO10}
Participants were requested to find the solution by performing an experiment, 
however, solving the problem by mathematical analysis is also recognized 
as alternative solution. 
The answer of the EPO10 problem is
\begin{equation}
U_\mathrm{v}=\epsilon.
\label{main_result}
\end{equation}
The goal of the theoretical calculations and the measurements is the same: 
understanding and \textit{pleasure of finding things out}.
And development of possible technical applications as a by-product.

\subsection{Introduction}
After big enough number of switchings, the voltage of the voltmeter
$U_\mathrm{v}$ 
does not depend on
$\mathcal{E}$, $C_\mathrm h$ and $C_\mathrm v$ 
and reaches a final value $\epsilon$.
Where could this independence find application?
-- In the measurement of small DC (direct current) voltages.
Imagine that $\mathcal E$ is a parasitic voltage of a 
precise DC.
This parasitic voltage is slowly changing with the time and
we address the old problem of the floating of the zero
of DC voltage amplifiers;
i.e. the voltmeter shows a small and slowly time dependent
voltage $\mathcal E(t)$ even if its input probes have the same 
electric potential.
The idea to overcome this trouble is very simple:
At short circuit at the input probes of the voltmeter
the parasitic voltage has to charge a capacitor in order to be remembered.
Than in the measuring phase, the capacitor has to be switched in
opposite direction in order parasitic voltage to be subtracted from 
the output of the amplifier and then the
voltmeter will show the measured voltage $\epsilon$.

In other words it is a method for measuring of DC voltages, 
whereby the switches automatically reset the device 
and cut the parasitic voltage $\mathcal{E}$. 
In such a way, the floating zero of the DC amplifiers is eliminated. 
Parasitic voltage $\mathcal{E}$ is compensated. 
This idea is patented~\cite{Goldberg1949} in 1949 
and it is realized in big number of contemporary  integrated 
circuits, auto-zero and chopper stabilized operational amplifiers.~\cite{Nolan2000,MT055,AD8571/AD8572/AD8574,AD8638/8639,MS2062,ADA4528} 
The authors of the patent have solved simple electrotechnical task with two capacitors, in which only a school physics knowledge, described in the present problem, is used.
After this general introduction we can return to our
school physics problem.

\subsection{Solution after big number of switchings}
\label{Solution after big number of switchings}
\begin{figure}[ht]
\includegraphics{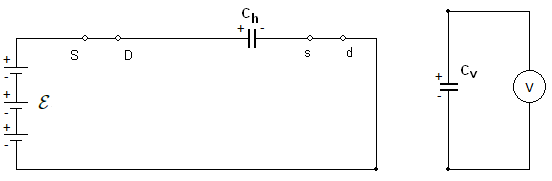}
\caption{In down position of the switches from FIG.~\ref{circuit} the battery $\mathcal{E}$ charges the capacitor $C_\mathrm{h}$ 
with a voltage $U_\mathrm{h}=\mathcal{E}$ (the sign can be precised), 
and the voltmeter shows voltage $U_\mathrm{v}$ 
on the capacitor $C_\mathrm{v}$.}
\label{circuit_down}
\end{figure}

\begin{figure}[ht]
\includegraphics{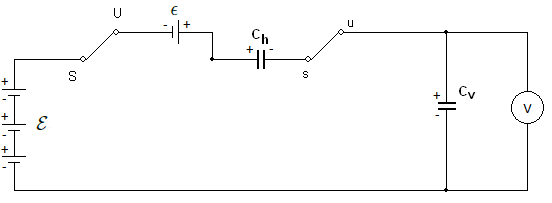}
\caption{In ``up position'' of the switches from FIG.~\ref{circuit} we have two batteries connected in series and two capacitors in series. The sum of the voltages of the batteries is equal to the sum of the voltages of the capacitors 
$\mathcal{E}+\epsilon=U_\mathrm{h}+U_\mathrm{v}$.
According to FIG.~\ref{circuit_down}
we have $U_\mathrm{h}=\mathcal{E}$ and consequently
we arrive at $U_\mathrm{v}=\epsilon$.
In other words, the problem of the EPO10 can be solved
only by carefully drawing of circuits.
The principle of the voltage compensation 
is illustrated in FIG.~\ref{battery_plus_plus}.}
\label{circuit_up}
\end{figure}

Regardless of the charge of the capacitor at the beginning, 
after a big number of switchings, 
a constant voltage $U_\mathrm{v}$ is established on the capacitor $C_\mathrm{v}$. 
This voltage is measured by a voltmeter. 
Let us begin the analysis of the scheme (circuit), 
when the switches are in down position (D and d). 
For clarity, the scheme is redrawn on FIG.~\ref{circuit_down} 
and the unnecessary details are removed.
In this case, the capacitor $C_\mathrm{h}$ is charged by the battery $\mathcal{E}$ to a voltage $U_\mathrm{h}=\mathcal{E}$, but the voltage of the capacitor $C_\mathrm{v}$ is $U_\mathrm{v}$. 
Let us ignore the slow discharge of $C_\mathrm{v}$ through the voltmeter. 
After one switching of the switch in up position (U and u) the charge of the capacitors is not changed, but now, 
as it is shown on FIG.~\ref{circuit_up}, 
we have two batteries and two capacitors connected in series. 
The sum of the voltages of the batteries is equal to the sum of the voltages on the capacitors 
$\mathcal{E}+\epsilon=U_\mathrm{h}+U_\mathrm{v}$. 
But as it was shown earlier $U_\mathrm{h}=\mathcal{E}$ 
and for the voltage of the capacitor $C_\mathrm{v}$ 
remains $U_\mathrm{v}=\epsilon$.
In such a way, the main result of the EPO10 
\Eqref{main_result} can be derived by purely theoretical consideration
of two circuits without making measurements.
This alternative method is also a highly appreciated solution.

This way, thanks to the switches, 
the voltage $\mathcal{E}$ is compensated 
and the voltmeter from FIG.~\ref{circuit} finally measures the voltage $\epsilon$, 
without being directly connected to the poles of the battery giving this voltage. 
This is the main detail of the described method for measuring of small
signals which are in some sense not directly connected 
to the voltmeter but always trough some parasitic voltages.
This result is not dependent on the ratio 
$C_\mathrm{v}/C_\mathrm{h}$, 
only the number of switchings should be significantly bigger than this ratio. 
In our case 
$C_\mathrm{v}/C_\mathrm{h}
=4700\;\mu\mathrm{F}/470\;\mu\mathrm{F}=10$, 
and for example, an acceptable accuracy is reached after, 
let say 20 switchings.

The DC amplifiers, which are used for measuring small voltages, 
''suffer'' from the so called ''floating zero'', 
the same way as the examined voltage $\epsilon$ 
is added to the parasitic voltage $\mathcal{E}$. 
That parasitic voltage can be eliminated by a system of switches and capacitors, very similar to the problem which we are solving. 
During the preparation of the EPO10 task, the amplifier was removed,
but the system of switches and capacitors, performing the 
self-resetting~\cite{Nolan2000,MT055} was left unchanged.
As we mentioned, the device is patented 74 years ago 
and is still one of the most frequently cited USA patents 
in the field of electronics~\cite{Goldberg1949}.

\subsection{Final remarks}

The general purpose of the Experimental-Physics Olympiad (EPO)
is to contribute to the education of experimental physics.
All EPOs problems where especially created for the Olympiads.
They where 
related to fundamental physics 
(for example, simple experimental set-ups for measurement of fundamental constants)
or technical patents used in contemporary devices.
\cite{EPO1,EPO2,EPO3,EPO4,EPO5,EPO6,EPO7,EPO8,EPO9}.
Some experimental set-ups given to students during EPOs are described
in great detail in Refs.~\cite{EJP1,EJP2,EJP3}.
The electronics related to the functioning of set-ups 
is described in specialized publications in engineering 
journals \cite{e1,e2,e3}.
The present problem is a popular version of technical notes
given by Analog Devices~\cite{Nolan2000,MT055}.
In addition this EPO10 task is a cover version of the EPO1 task.
Let us summarize what we have observed during these 10 years.
When the problem is given at our Olympiads 
there are always several students capable to understand and solve 
the problem to great detail which is very encouraging.
However, everybody experienced difficulty 
to guest what could be the technical application,
which indicate the need to complement 
the education in physics.

\clearpage

\section{Tasks S. Getting to know the voltage source}

\begin{enumerate}

\item 1.625~V, 1.630~V, 1.625~V, 1.634~V.

\item  -+-+: 3.27~V, -++-: 0.002~V.

\item 1.630~V.

\item 4.9~V.

\item Capacitor charge.

\section{Tasks M. Charging and discharging of capacitors}

\item The results from the measurements are presented 
in Table~\ref{table_C_discharge}.
\begin{table}[ht]
\begin{tabular}{ r  r  r  r}
\tableline \hline
i & $\qquad t_\mathrm{i}$ [s] & $\qquad U_\mathrm{i}$ [mV] & 
$\qquad  -\ln(U_\mathrm{i}/U_0)$\\
\tableline 
0	&	0	&	1623	&	0	\\
1	&	30	&	1537	&	0.0544	\\
2	&	60	&	1444	&	0.1169	\\
3	&	90	&	1365	&	0.1731	\\
4	&	120	&	1280	&	0.2374	\\
5	&	150	&	1213	&	0.2912	\\
6	&	180	&	1140	&	0.3532	\\
7	&	210	&	1080	&	0.4073	\\
8	&	240	&	1014	&	0.4704	\\
9	&	270	&	960	&	0.5251	\\
10	&	300	&	902	&	0.5874	\\
11	&	330	&	854	&	0.6421	\\
12	&	360	&	801	&	0.7062	\\
13	&	390	&	760	&	0.7587	\\
14	&	420	&	714	&	0.8211	\\
15	&	450	&	675	&	0.8773	\\
16	&	480	&	636	&	0.9368	\\
17	&	510	&	602	&	0.9918	\\
18	&	540	&	565	&	1.0552	\\
19	&	570	&	536	&	1.1079	\\
20	&	600	&	503	&	1.1714	\\
\tableline \hline
\end{tabular}
\caption{Results from the measurements of the voltage 
of the capacitor as a function of time.}
\label{table_C_discharge}
\end{table}

\item \Fref{U-t}.
\begin{figure}[ht]
\includegraphics[scale=0.6]{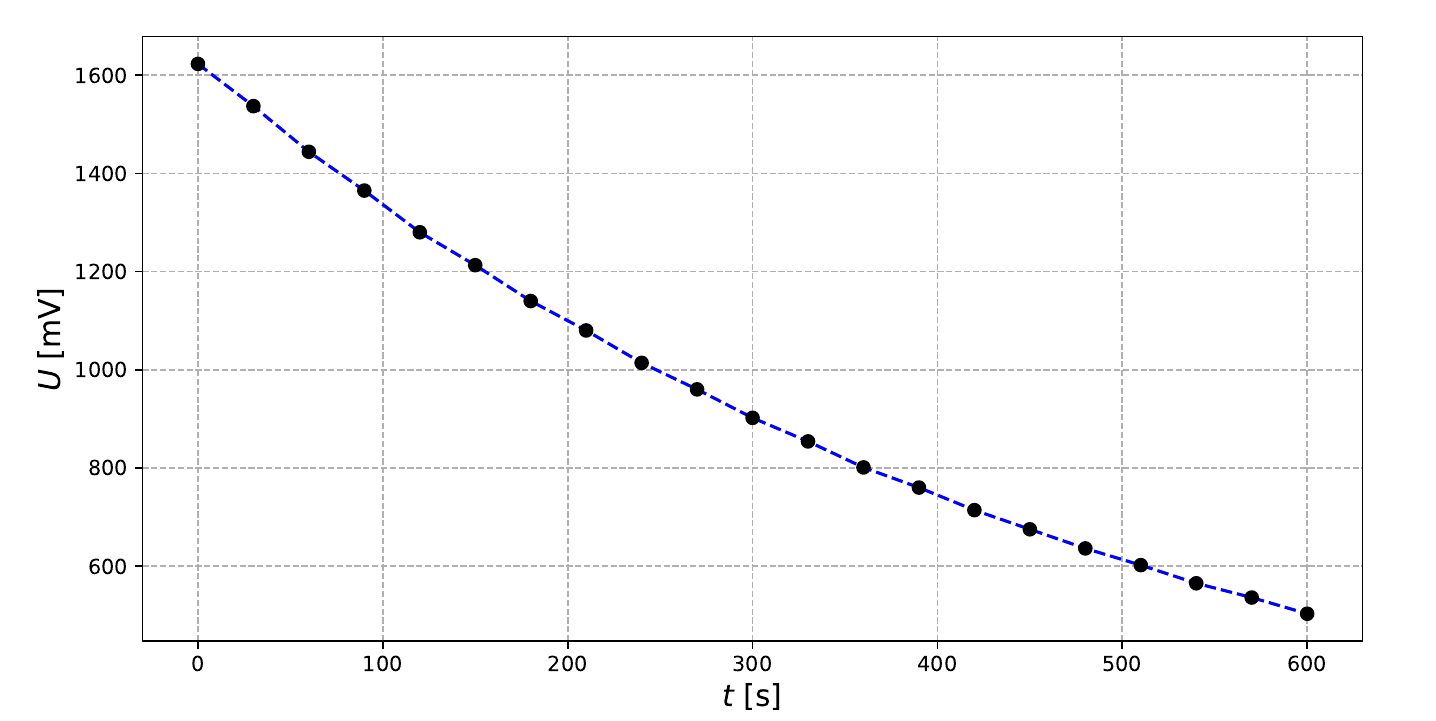}
\caption{Time-dependence  of the capacitor voltage $U(t)$ with the measurements from Table~\ref{table_C_discharge}.}
\label{U-t}
\end{figure}

\item Table~\ref{table_C_discharge}, last column (column 4).

\item \Fref{ln(U)-t}. The quantity is frequency [Hz] or reciprocal time [s], which is evident from the figure since the abscissa dimensionality is time and the ordinate is dimensionless. 
\begin{figure}[ht]
\includegraphics[scale=0.7]{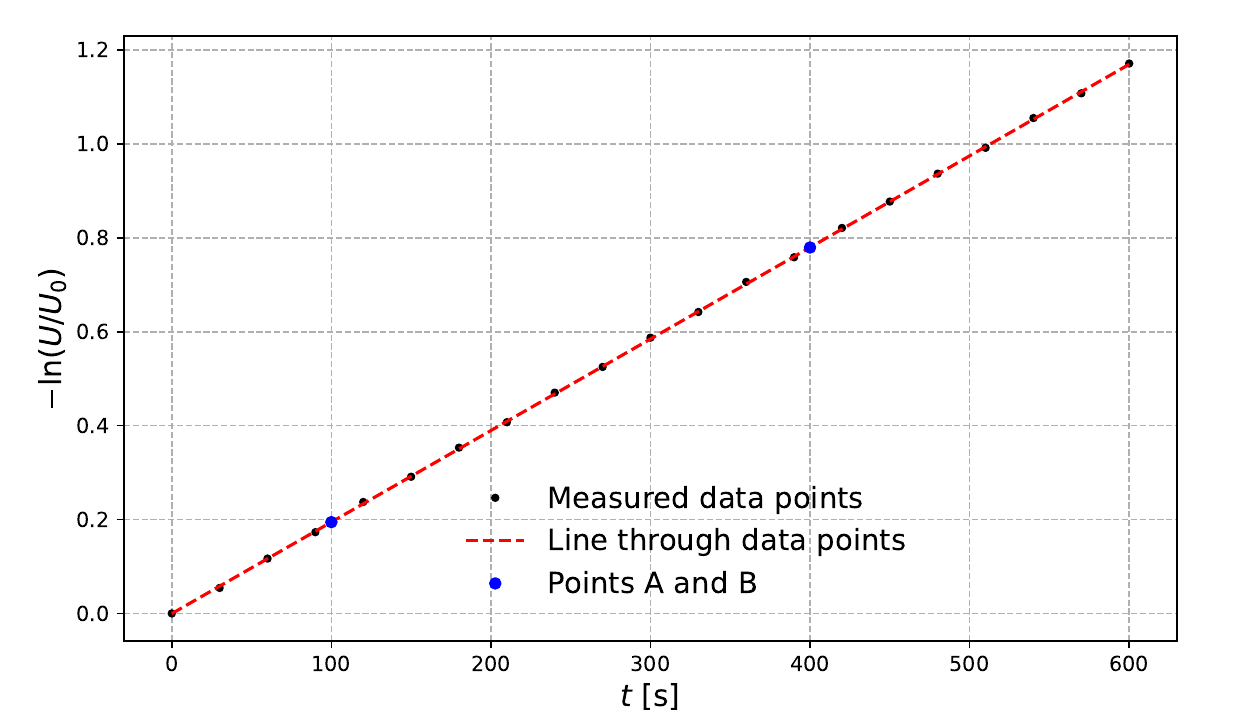}
\caption{
Time-dependence of the logarithm of dimensionless capacitor voltage 
$-\ln[U(t)/U_0]$.
Points A and B are chosen to be on the approximating straight line, while not all experimental data points lie on the line.
Due to the precise measurement in this case, most experimental points are very close to the line, which points to small experimental error.}
\label{ln(U)-t}
\end{figure}

\item Points A and B in \Fref{ln(U)-t}, for which
$t_\mathrm{A}=100~\Omega$ and $t_\mathrm{B}=400~\Omega$ and the ordinate values
$-\ln \left[U(t)/U(0)\right]_\mathrm{A}=0.1945$ and 
$-\ln \left[U(t)/U(0)\right]_\mathrm{B}=0.7795$.
Therefore
\be
a = \frac{1}{\tau} = 1.95 \times 10^{-3}~[\mathrm{s^{-1} \equiv Hz}].
\nn
\ee

\item $\tau = 1/a \approx 512.8$~s.
The dimension is time because it is reciprocal to the slope of the line $a$.

\item $R = 0.97~\mathrm{M}\Omega$.

\item $C = \tau/R = 528.68 \approx 529~\mu$F.

\item Taking nominal value written on the capacitor $C_\mathrm{n}=470~\mu$F.

\item $\Delta C = | C - C_\mathrm{n} |/C_\mathrm{n} = 0.126 \times 100 = 12.6$\%.
More accurate value is the measured one $C$ because of the precise measurement evident from the excellent match of experimental points and approximating line in \Fref{ln(U)-t}.

\section{Tasks L. The experiment with the given circuit}

\item Circuit connection.

\item Capacitor polarity check.

\item First measurements of $U_\mathrm{v}$, matches \#1 from Table~\ref{solution_table}.

\item First measurements of $U_\mathrm{h}$, matches \#1 from Table~\ref{solution_table}.

\item Table~\ref{solution_table}.

\begin{table}[ht]
\begin{tabular}{ c  r  r  r  r}
\tableline \tableline \tableline
\# &$\quad \mathcal{E}$ (V) & $\quad \epsilon$ (V) 
& $\quad U_\mathrm{v}$ (V) & $\quad U_\mathrm{h} (V) $ \\
\tableline \hline
1&4.910 & 1.632 & 1.622 & 4.95 \\
2&3.270 & 1.632 & 1.626 & 3.31 \\
3&1.632 & 1.632 & 1.628 & 1.70 \\
4&0.0 & 1.632 & 1.628 & 0.04 \\
\tableline \hline
\end{tabular}
\caption {Solution of large enough fast switching, 1 multimeter was used and therefore $U_\mathrm{h}$ and $U_\mathrm{v}$ were measured one after another.
One can see that $U_\mathrm{v}\approx\epsilon$
and
$U_\mathrm{h}\approx\mathcal E$.
}
\label{solution_table}
\end{table}

\item Table~\ref{solution_table} rows after \#1.

\item Analyzing Table~\ref{solution_table} we observe that
within experimental accuracy $U_\mathrm{v}=\epsilon$.

\item $\mathcal{E}+\epsilon=U_\mathrm{h}+U_\mathrm{v}$.

\section{Tasks L and XL. Possible practical application of the solved problem?}

\item The equation  $U_\mathrm{v}=\epsilon$ is actually an exact result 
on which are based auto-zero DC
amplifiers in which parasitic offset voltage 
$\mathcal{E}$ is compensated.
This item related to possible technical applications 
was not solved by any participant.
To solve a problem and to estimate where it can be applied are very different tasks confer Refs.~\cite{Goldberg1949,Nolan2000,MT055,AD8571/AD8572/AD8574,AD8638/8639,MS2062,ADA4528}.

\section{Tasks L and XL. Theoretical problems related to the experiment}

\item The voltage of the capacitor can be expressed by the charge of the capacitor $U(t)=Q(t)/C$.
The current trough the voltmeter is just the time derivative of this charge
$I(t)=\mathrm{d}Q/\mathrm{d}t$.
But the voltage of the voltmeter is equal to the voltage of the capacitor,
and according the Ohm law $U(t)=RI(t)$. 
Substituting here the current by the charge derivative we obtain
$Q(t)/C=-R\mathrm{d}Q/\mathrm{d}t$;
the sign minus describes that charge decreases.
Now introducing $\tau\equiv RC$
this equation can be rewritten as
$\mathrm{d}Q/\mathrm{d}t=-Q/\tau$.
One can easily check that this equation has the solution
$Q(t)=Q(t=0)\exp(t/\tau)$.
Substitution in the formulas for current and voltage
gives $I(t)=I_0\exp(t/\tau)$ and $U(t)=U_0\exp(t/\tau)$,
where $U_0=Q_0/C$ and $I_0=U_0/RC$.

\item The time constant  $\tau$ can be easily determined without calculating the slope of the line $a$.
Exponential voltage decay $U(t)=U(0)\exp(-t/\tau)$ 
in time means that for each passed time equal to the time constant, the voltage decreases e times.
\Fref{ln(U)-t} ordinate is in negative ln scale meaning that $e$ times decrease is actually equal to 1.
Therefore for $t=\tau$ we have $-\ln(U_\tau/U_0)=1$.
Now, looking at Table~\ref{solution_table}
in the last column we search for the nearest value of 1, which in our case is 0.9918 for i=17 and $t_\mathrm{i}=\tau=510$~s.
Comparing with the more accurate method 
$\Delta \tau \approx 3.6$\%
which is an excellent accuracy given the so much less effort in achieving it.
Moreover, this quick method could be significantly improved by taking measurements of the voltage in evenly spaced time intervals 
(like each 30 seconds)
but taking measurements of the time intervals $t_\mathrm{i}$ for the  values of the voltage $U_\mathrm{i}$ for which
\be
-\ln(U_\mathrm{i}/U_0)=0.25,0.5,0.75,1.0, \dots
\nn
\ee
$U_0$ is known before the experiment, so the values of $U_\mathrm{i}$ are preliminary calculated and tabulated.
In this way several values for $\tau$ will be obtained
\be
\tau_\mathrm{i} = 4.0, 2.0, 0.75, 1.0, \dots \times t_\mathrm{i}
\nn
\ee
and either average or median can be taken.
In short we have to calculate $U(0)/2.718$ to to wait 
when $U(\tau)=U(0)/2.718$.
This method is however nonapplicable for very big capacitors
and good voltmeters,
for example for $C=4700~\mu\mathrm{F}$ and 
$R=10\,\mathrm{M}\Omega$ we have
$\tau=RC=48\,$ks which is comparable with 
24*3600=86.4~ks.
An alternative experimental method is described in 
subsection~\Ref{Afterglow}..

\item 
The solution is given in 
subsection~\Ref{Solution after big number of switchings}.

\end{enumerate}

Now when the experimental problem is analyzed 
one can easily understand the work of
auto-zero and chopper stabilized operational 
amplifiers ~\cite{Nolan2000,MT055,MS2062}
and implementation of the idea to use
charged capacitor to compensate drift of zeros 
in the description of many operational 
amplifiers~\cite{AD8571/AD8572/AD8574,AD8638/8639,ADA4528}.
Only now one can read and understand the initial idea in the 
patent~\cite{Goldberg1949}.
We will be glad if in the future some of EPO participants 
would have significant recognized inventions.

\section{Afterglow}
\label{Afterglow}

\begin{figure}[ht]
\includegraphics[scale=0.5]{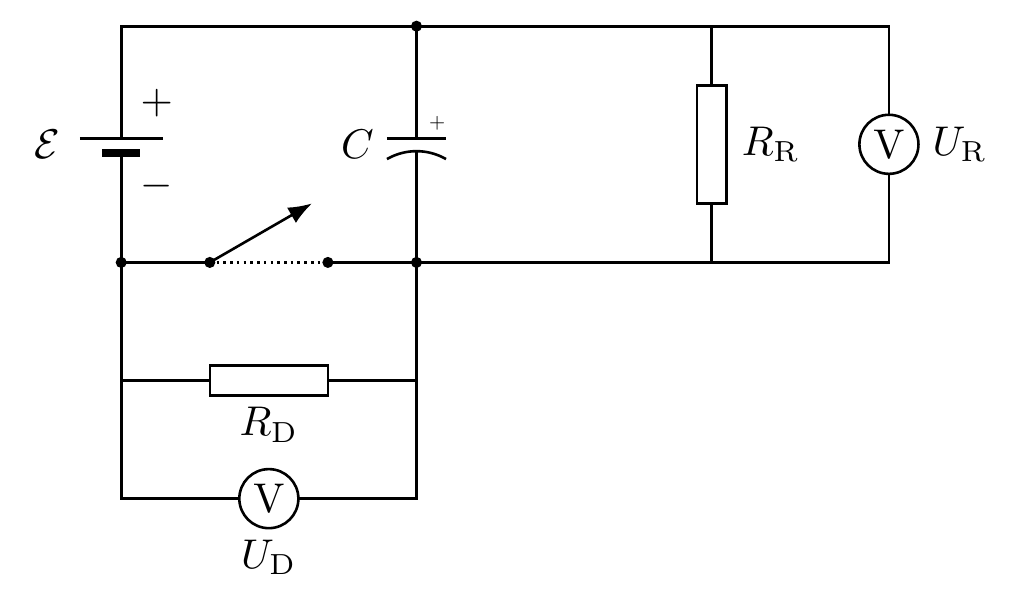}
\caption{Measuring of capacitance $C$ of a big electrolytic capacitor using two voltmeters with internal resistance $R_\mathrm{R}=R_\mathrm{D}=R$.
In the beginning the switch is in down position and the battery 
with electromotive force $\mathcal{E}$ charge the the capacitor with voltage
$U_\mathrm{R}(0)=\mathcal E$.
The down voltmeter connect equi-potential points and 
$U_\mathrm{D}(0)=0$.
In $t=0$ the switch is switched in up position add the capacitor 
is slowly discharging trough two parallel resistors.
The time-constant is $\tau=CR/2$.
In every moment $U_\mathrm{D}(t)+U_\mathrm{R}(t)=\mathcal E$. 
After long time $t\gg\tau$ we have a voltage divider 
$U_\mathrm{D}(\infty)=U_\mathrm{R}(\infty)=\mathcal E/2$. 
}
\label{e-consv}
\end{figure}

The circuit for measuring of big capacitors is depicted at
FIG.~\ref{e-consv}.
Theoretical analysis of this circuit gives that
\be
U_\mathrm{D}(t)=(1-\mathrm{e}^{-t/\tau})\mathcal E/2,\qquad
U_\mathrm{R}(t)=(1+\mathrm{e}^{-t/\tau})\mathcal E/2.
\ee
Or for small times $t\ll\tau$ using that $\exp x\approx 1+x$
\be
U_\mathrm{D}(t)\approx \frac{t}{\tau}\,\mathcal E/2
\ee
Substituting here $\tau=RC/2$
we obtain
\be
C\approx\frac{t}{R}\,\frac{\mathcal{E}}{U_\mathrm{D}(t)}.
\ee
For example 
using 3 AAA batteries and
starting with $U_\mathrm{R}(0)=\mathcal{E}=4.95\,$V
the voltage reaches 
$U_\mathrm{D}(t)=495\,\mathrm{mV}=\mathcal{E}/10$
(we used 2000~mV range)
after $t=7\,\mathrm{min}+17\,\mathrm{ses}= 437\,$s.
In such a way we determine 
$\tau=4370\, \mathrm{sec}=$~1~hour+12~min+50~sec.
And finally for $R=1\,\mathrm{M}\Omega$ we determine 
$C=4370\,\mu$F. Which is  less 10\% from the nominal value
$C=4700\,\mu$F
\be
\frac{\Delta C}{C}=\frac{430}{4700}<0.1.
\ee
Using this method is possible to determine 
big capacitance within 10 minutes and without loss of precision.

\acknowledgments
The present EPO10 is continuation of EPO1 problem
with adaptation and introduction of new content.
That is why we would like to 
thank Stojan Manolev, Vassil Jordanov and Peter Peshev
for the collaboration and co-authorship during EPO1 back in 2014.
Discussions with Peter Todorov, George Bakos and Nikola Serafimov 
are highly appreciated.
The work of the EPO10 absolute champion Maximilian Jelacic is scanned and attached as a separate pdf file named ``EPO10\_Abs-Champ.pdf'' to the current arXiv version.



\begin{thebibliography}{9}
%
\bibitem{Goldberg1949}
  Edwin Goldberg and Jules Lehmann,
  \emph{Stabilised direct current amplifier},
  U.S. Patent 2,684,999 (1949); \\
 \url{http://www.philbrickarchive.org/us2684999_chopper.pdf}

\bibitem{Nolan2000}
  Eric Nolan,
  \emph{Demystifying Auto-Zero Amplifiers—Part 1.
  They essentially eliminate offset, drift, and 1/f noise. How do they work? Is there a downside?},
  Analog Dialogue \textbf{34-2} (2000) pp. 1-3, \textbf{Fig.~1};
  \textit{ibidem} 34-3 pp. 1-2; \\
 \url{http://www.analog.com/library/analogDialogue/archives/34-02/demystify/demystify.pdf}

\bibitem{MT055}
  \emph{Chopper Stabilized (Auto-Zero) Precision Op Amps},
  Analog Devices, MT-055 TUTORIAL (2009) pp. 1-6, Fig.~2; \\
  \url{http://www.analog.com/static/imported-files/tutorials/MT-055.pdf}

\bibitem{MS2062}
  Reza Moghimi,
  \emph{To Chop or Auto-Zero: That Is the Question},
  Analog Devices, MS-2062 Technical Article (2011)  pp. 1-6, Figs. 1, 2; 
 \url{http://www.analog.com/static/imported-files/tech_articles/MS-2062.pdf}

\bibitem{AD8571/AD8572/AD8574}
  \emph{Zero-Drift, Single-Supply, Rail-to-Rail Input/Output Operational Amplifier},
  Analog Devices, Datasheet AD8638/AD8639, (1999-2011); 
 \url{http://www.analog.com/static/imported-files/data_sheets/AD8571_8572_8574.pdf}

\bibitem{AD8638/8639}
  \emph{16 V Auto-Zero, Rail-to-Rail Output Operational Amplifiers},
  Analog Devices, Datasheet AD8571/AD8572/AD8574, (2007-2010); 
  \url{http://www.analog.com/static/imported-files/data_sheets/AD8638_8639.pdf}
 
\bibitem{ADA4528}
  \emph{Precision, Ultralow Noise, RRIO, Zero-Drift Op Amp},
  Analog Devices, Datasheet ADA4528-1/ADA4528-2, (2011-2012); \\
 \url{http://www.analog.com/static/imported-files/data_sheets/ADA4528-1_4528-2.pdf}

\bibitem{EPO1}
V.~G.~Yordanov, P.~V.~Peshev, S.~G.~Manolev, T.~M.~Mishonov,
``Charging of capacitors with double switch. The principle of operation of auto-zero and chopper-stabilized DC amplifiers'',
arXiv:1511.04328 [physics.ed-ph]

\bibitem{EPO2}
V.~N.~Gourev, S.~G.~Manolev, V.~G.~Yordanov, T.~M.~Mishonov,
``Measuring Plank constant with colour LEDs and compact disk'',
arXiv:1602.06114 [physics.ed-ph].

\bibitem{EPO3}
S.~G.~Manolev, V.~G.~Yordanov, N.~N.~Tomchev, T.~ M.~Mishonov,
``Volt-Ampere characteristic of "black box" with a negative resistance'',
arXiv:1602.08090 [physics.ed-ph].

\bibitem{EPO4}
V.~G.~Yordanov, V.~N.~Gourev, S.~G.~Manolev, A.~M.~Varonov, T.~M.~Mishonov,
``Measuring the speed of light with electric and magnetic pendulum'',
arXiv:1605.00493 [physics.ed-ph].

\bibitem{EPO5}
T.~M.~Mishonov, E.~G.~Petkov, A.~A.~Stefanov, A~P.~Petkov, I.~M.~Dimitrova,
S.~ G.~Manolev, S.~I.~Ilieva, A.~M.~Varonov,
``Measurement of the Boltzmann constant by Einstein. Problem of the 5-th Experimental Physics Olympiad. Sofia 9 December 2017''
arXiv:1801.00022v4 [physics.ed-ph].

\bibitem{EPO6}
T.~M.~Mishonov, E.~G.~Petkov, A.~A.~Stefanov, 
A.~P.~Petkov, V.~I.~Danchev, Z.~O.~Abdrahim, Z.~ D.~Dimitrov, 
I.~M.~Dimitrova, R.~Popeski-Dimovski, M.~Poposka, 
S.~Nikoli\'c, S.~Miti\'c, R.~Rosenauer, F.~Schwarzfischer, 
V.~N.~Gourev, V.~G.~Yordanov, A.~M.~Varonov
``Measurement of the electron charge $q_e$ using Schottky noise. 
Problem of the 6-th Experimental Physics Olympiad. Sofia 8 December 2018'',
arXiv:1703.05224v2 [physics.ed-ph].

\bibitem{EPO7}
T.~M.~Mishonov, R.~Popeski-Dimovsk, L.~Velkoska, I.~M.~Dimitrova,
V.~N.~Gourev, A.~P.~Petkov, E.~G.~Petkov, A.~M.~Varonov,
``The Day of the Inductance. Problem of the 7-th Experimental Physics Olympiad, Skopje, 7 December 2019'',
arXiv:1912.07368 [physics.ed-ph].

\bibitem{EPO8}
T.~M.~Mishonov, A.~P.~Petkov, M.~Andreoni, E.~G.~Petkov, A.~M.~Varonov, I.~M~ Dimitrova, L.~Velkoska, R.~Popeski-Dimovski,
`Problem of the 8th Experimental Physics Olympiad, Skopje, 8 May 2021 Determination of Planck constant by LED`'',
arXiv:2106.01337 [physics.ed-ph].

\bibitem{EPO9}
T.~M.~Mishonov, N.~S.~Serafimov, E.~G.~Petkov, A.~M.~Varonov,
``Set-up for observation thermal voltage noise and determination of absolute temperature and Boltzmann constant'',
arXiv:2205.06609 [physics.ed-ph].

\bibitem{EJP1} 
T.~M.~Mishonov, A.~M.~Varonov, D.~D.~Maksimovski, S.~G.~Manolev, 
V.~N.~Gourev and V.~G.~Yordanov,
``An undergraduate laboratory experiment for measuring $\varepsilon_0$, $\mu_0$ 
and speed of light $c$ with do-it-yourself catastrophe machines: 
electrostatic and magnetostatic pendula'',
Eur.~J.~Phys. \textbf{38}, 025203 (2017).

\bibitem{EJP2} 
T.~M.~Mishonov, V.~N.~Gourev, I.~M.~Dimitrova, N.~S.~Serafimov, 
A.~A.~Stefanov, E.~G.~Petkov and A.~M.~Varonov
``Determination of the Boltzmann constant by the equipartition theorem for capacitors'',
Eur. J. Phys. \textbf{40}, 035102 (2019).

\bibitem{EJP3}
T.~M.~Mishonov, E.~G.~Petkov, N.~Zh.~Mihailova, A.~A.~Stefanov, 
I.~M.~Dimitrova, V.~N.~Gourev, N.~S.~Serafimov, V.~I.~Danchev, and A.~M.~Varonov,
``Simple do-it-yourself experimental set-up for electron charge $q_e$ measurement'',
Eur.~J.~Phys. \textbf{39}, 065202 (2018).

\bibitem{e1}
T.~M.~Mishonov, V.~I.~Danchev, E.~G.~Petkov, V.~N.~Gourev, I.~M.~Dimitrova, N.~S.~Seraﬁmov, A.~A.~Stefanov and A.~M.~Varonov,
``Master equation for operational ampliﬁers: stability of negative
differential converters, crossover frequency and pass-bandwidth'',
J. Phys. Commun.~\textbf{3}, 035004 (2019).

\bibitem{e2}
T.~Mifune, T.~M.~Mishonov, N.~S.~Serafimov, I.~M.~Dimitrova,
R.~Popeski-Dimovski, L.~Velkoska, E.~G.~Petkov, A.~M.~Varonov, A.~Barone,
``Tunable high-Q resonator by general impedance converter'',
Rev. Sci. Instrum.~\textbf{92}, 025123 (2021).

\bibitem{e3} 
T.~M.~Mishonov, E.~G.~Petkov, I.~M.~Dimitrova, N.~S.~Serafimov and A.~M.~Varonov,
``Probability distribution function of crossover frequency of operational amplifiers'',
Measurement~\textbf{179}, 109509 (2021).

\bibitem{IYL2015}
  \emph{Second Experimental Physics Olympiad: The Day of the Photon in the International Year of Light}, Sofia, 25 April (2015); \\
 \url{http://www.light2015.org/Home.html}

\bibitem{EPO_IYL2015}
  \emph{Second Experimental Physics Olympiad: The Day of the Photon in the International Year of Light}, Sofia, 25 April (2015); 
 \url{http://www.light2015.org/Home/Event-Programme/2015/Competition/Bulgaria-Second-Experimental-Physics-Olympiad--25-April-2015-in-Sofia.-The-Day-of-the-Photon-in-the-International-Year-of-the-Light.html}

\end{thebibliography}
\end{document}